\magnification=1200
\parindent=1.5truecm
\baselineskip 0.6cm
\centerline{\bf NONLINEAR FEEDBACK OSCILLATIONS}
\centerline{\bf IN RESONANT TUNNELING THROUGH DOUBLE BARRIERS}
\vskip 1.5cm
\centerline{\bf C. Presilla}
\centerline{ Scuola del Dottorato di Ricerca in Fisica
 dell' Universit\'a ``La Sapienza'', Roma, I 00185}
\centerline{ and}
\centerline{ Dipartimento di Fisica dell' Universit\'a, Perugia, I 06100}
\centerline{\bf G. Jona-Lasinio}
\centerline{ Dipartimento di Fisica dell' Universit\'a ``La Sapienza'', Roma, 
 I 00185}
\centerline{\bf and}
\centerline{\bf F. Capasso}  
\centerline{ AT\&T Bell Laboratories, 600 Mountain avenue,
 Murray Hill, N J 07974}
\vskip 3cm
\baselineskip 0.8cm
\centerline{\sl Abstract}
 We analyze the dynamical evolution of the
 resonant tunneling of an ensemble of electrons through a double barrier in
 the presence of the self-consistent potential created by the charge
 accumulation in the well. 
 The intrinsic nonlinearity of the transmission process is shown 
 to lead to oscillations of the stored charge and of the transmitted 
 and reflected fluxes.
 The dependence on the electrostatic feedback induced by the self-consistent
 potential and on the energy width of the incident distribution is discussed.
\vskip 1.0cm
\leftline{P.A.C.S. numbers: 03.65.-w, 73.40Gk }
\vfill \eject
 
In recent years there has been renewed interest in the phenomenon of
 resonant tunneling (RT) through double barriers.
 The unique capabilities of molecular beam epitaxy make it possible
 to investigate fundamental questions on RT through simple man-made 
 potentials by controlling the barrier and well parameters (e.g. height,
 thickness or barrier phase area) down to the atomic scale$^{[1]}$.
 
In this paper we investigate the dynamics of RT of ballistic electrons
 in the presence of the potential created by the charge trapped 
 within the barriers$^{[2]}$.
 This problem is interesting not only from a technological point of view
 but also as a test of quantum mechanical non-equilibrium situations in
 which many particles are involved.

The model we propose tries to describe the following situation.
 A bunch of electrons is created within a contact layer and launched towards
 embedded semiconductor layers forming a double barrier potential.
 The charge dynamically trapped by the resonance will produce a reaction
 field which modifies the time evolution of the system.
 An exact treatment of such a problem looks very complicated.
 We assume a decoupling between the longitudinal (in the direction $x$ of 
 motion perpendicular to the double barrier) and transversal degrees
 of freedom. 
 This is a common assumption in  treating tunneling phenomena.
 It makes the problem one-dimensional and allows the 
 following factorization of the wave function:
$$
\Psi({\bf x}_1,{\bf x}_2,...,{\bf x}_N;t)=
\Phi(x_1,x_2,...,x_N;t)~
\Omega(y_1,z_1,y_2,z_2,...,y_N,z_N;t)
\eqno (1)
$$
 where ${\bf x} \equiv x,y,z$ are the coordinates of each electron
 and $t$ the time.
 We remark that antisymmetry of $\Psi$ is established if we take, for
 example, $\Phi$ symmetric and $\Omega$ antisymmetric in their arguments.
 The experimental set-up we have in mind puts ideally all the electrons
 in the same high-energy longitudinal state while the transversal degrees
 of freedom are essentially decoupled$^{[1]}$.
 Therefore the choice (1) with $\Phi$ symmetric is the only possibility.
 In other words the transversal degrees of freedom assure the respect of 
 Pauli principle. 

Finally we assume the electrons in the bunch at the initial time
 uncorrelated which corresponds to a choice of $\Phi$ as a product
 state of single-particle states $\psi(x,0)$.
 At this point theorem 5.7 of Ref. [3] guarantees in the mean field 
 approximation 
 (which is reasonable due to the large number of electrons involved)
 that the state $\Phi$ remains a product state during its evolution
 and allows us to write the following self-consistent equation for
 $\psi(x,t)$:
$$
i \hbar {\partial \over \partial t} \psi (x,t) =
 \biggl [ - {\hbar^2 \over 2 m} {\partial^2 \over \partial x^2}
 + V(x) + \int W(t,t';x,x') |\psi(x',t')|^2 dt' dx'
 \biggr ] \psi(x,t)
\eqno (2)
$$

The external potential, $V(x)$, is assumed, as customary, to be a step
 function (no electric field is applied):
$$ 
V(x)=V_0 [\theta(x-a) \theta(b-x) + \theta(x-c) \theta(d-x)]
\eqno (3)
$$
  with $a<b<c<d$ and where $\theta(x)$ is the Heaviside function.  
 The nucleus $W(t,t';x,x')$ is modelled assuming that memory
 effects can be neglected, i.e. $W\propto \delta (t-t')$.
 We represent the global repulsive feedback effect,
 induced by the charge localized in the well, by a shift of 
 the bottom of the well to a higher energy, $V_Q(t)$,
 proportional to the charge, i.e.:
$$
\int W(t,t';x,x') |\psi(x',t')|^2 dt' dx' 
\equiv \alpha V_0 {Q(t) \over Q_0} \theta(x-b) \theta(c-x)
\eqno (4)
$$
 $Q(t)$ is the charge localized in the well at time $t$ and 
 $Q_0$ is a normalization charge which depends on
 the shape of the initial state, assumed localized around $x_0$:
$$
 Q(t) \equiv \int_b^c dx' |\psi(x',t)|^2 ~~~~~~~~~~
 Q_0\equiv \int_{x_0 -(c-b)/2}^{x_0 +(c-b)/2} dx' |\psi(x',0)|^2
\eqno (5)
$$
 $Q_0$ introduces an artificial dependence of the Eq. (2) on the
 initial condition. We have used this parameterization to make
 the comparison of different numerical simulations easier.
 The parameter $\alpha$ in Eq. (4) can be varied to reproduce 
 phenomenologically the response of the medium to the
 charge trapped in the well and the characteristics of the electron
 bunch, e.g. its areal density.

The 1-particle state which is the initial condition in our mean field 
 equation (Eq. 2) has been chosen to be a gaussian shaped
 superposition of plane waves with mean momentum $\hbar k_0$:
$$
 \psi (x,0) = {1 \over \sqrt{\sigma \sqrt{\pi} } }
 \exp{\biggl [- {1 \over 2} \biggl ( {x-x_0 \over \sigma} \biggr )^2 
 + i k_0 x \biggr ]}
\eqno (6)
$$
 with energy spread
 (energy full width at half maximum of the square modulus of the
 Fourier transform of (6))
 $\Gamma_0= 2 \sqrt{\ln 2}~ \hbar^2 k_0 /  m \sigma$.
 $x_0$ is chosen so that at the initial time no appreciable charge
 sits in the well, i.e. $Q(0)=0$.

The solution of the differential equation (2-5) with the initial condition 
 (6) has been achieved by a numerical integration on a two-dimensional
 lattice$^{[4]}$.
 Assuming for the barrier and well widths the values 
 $b-a=d-c=20~a_0$ and $c-b=15~a_0$ ($a_0 \simeq 0.529~ \AA$ being the 
 Bohr radius) and for the barrier height $0.3~eV$ and using for $m$
 the free electron mass, the potential $V(x)$
 exhibits a single resonance in the transmission coefficient at energy 
 $E_R \simeq 0.15~eV$, the shape of which is well approximated by a lorentzian
 of full width $\Gamma_R \simeq 5~meV$.
 The choice of these parameters was a compromise
 between the requirement of standard technological values and 
 that of reasonable computation times.
 The electron mass was set to its free value to avoid the complications
 of a space variable effective mass.

The incoming state mean-energy has been chosen so as to satisfy
 the resonance condition $\hbar^2 k_0^2/2m = E_R$. 
 Finally, the normalized charge in the well, $Q(t)/Q_0$, has been plotted
 as a function of time.
 The results for different choices of the parameter $\alpha$ are shown 
 in Figs. 1-4 in the case of states with energy-spread larger,
 of the order of and smaller than the resonance width.

The Fig. 1 represents the scattering of a wave packet on 
 the fixed double barrier (linear Schr$\ddot o$dinger equation). 
 When the packet is energetically much wider than the resonance, 
 the building up and the decay of the charge are asymmetric,
 the decreasing following$^{[5]}$ the law
 $exp(-t/ \tau)$ with $\tau=\hbar / \Gamma_R$
 (decay of a lorentzian-shaped quantum state).
 On the other hand, for a wave packet narrower than the resonance, 
 the charge presents a symmetric behavior, like 
 the law $exp[-((t-t_0)/ \tau)^2]$, where $\tau=\sigma / v_0
 = 2 \sqrt{\ln 2}~ \hbar / \Gamma_0 $,
 $~t_0\simeq |(b+c)/2-x_0|/v_0$ and $v_0=\hbar k_0 /m$ (free evolution of
 a gaussian-shaped quantum state for negligible time-spreading). 
 The latter result is not surprising, since now the packet traverses
 the double barrier almost undistorted. 
 In the case of a state energy-spread comparable with
 the resonance width, the evolution of the trapped charge interpolates
 between these two extreme behaviors.  

When the non-linear term is effective, i.e. $\alpha \neq 0$, the 
 evolution of the trapped charge changes drastically and  
 oscillations can appear.
 This phenomenon has been qualitatively anticipated by Ricco and
 Azbel$^{[6]}$. 
 Their reasoning was very simple. 
 At the initial time no charge is present in the well, the electrons are
 moving towards the double barrier and the resonance condition
 is fulfilled.
 When some charge penetrates into the well, the modification of the
 potential destroys the resonance condition.
 As a consequence the quantity of trapped charge has a maximum 
 followed by a decrease.
 The resonance condition tends to be restored and a new cycle begins.
 However, as it will appear in the following, the nonlinearity makes the
 interpretation of the phenomenon significantly more complicated.
 For example, the conclusion by Ricco and Azbel that the above effect
 should be maximal for monochromatic states, is not correct.

A detailed analysis of Figs. 2-4 (that are a sample of our
 global numerical work) suggests the following observations.
 Oscillations are present, for appropriate values of the strength of the
 non-linear term, $\alpha$, only when the energy spread of the state
 is wider or comparable to the resonance width.
 No oscillations are seen for nearly monochromatic states. 
 When $\alpha$ increases, the oscillations, if present, tend to increase
 in number but decrease in amplitude.

To understand these results, let us first interpret the dependence
 of the intensity of the trapped charge as a function of the parameters
 $\alpha$ and $\sigma$.
 We simplify the question by considering
 a time-average of the charge dynamically present inside the well.
 Since during the time evolution $V_Q(t)$ and $Q(t)$ are related
 with each other by the Eq. (4), a similar relation has to hold between
 the relevant time averaged quantities denoted by $V_Q$ and $Q$.
 Let us suppose, now, that we have a time independent situation with the bottom
 of the well at level $V_Q$.
 As can be shown by explicit calculations, the
 charge $Q$ present in the well is a fraction, $\gamma$, of the asymptotically
 transmitted charge $Q_T$$^{[5]}$:
$$
Q_{T}(V_Q)= \int_{- \infty}^{+ \infty} dk~ |\widetilde \psi(k,0)|^2~ 
|t_{V_Q}(k)|^2
\eqno (7)
$$
 where $\widetilde \psi(k,0)$ is the Fourier transform of (5) and
 $|t_{V_Q}(k)|^2$ is the transmission coefficient of the depicted potential.
 The time-average of Eq. (4) can be combined with Eq. (7)
 to obtain a self consistent relationship for $V_Q$ (or $Q$):
$$
{V_Q \over \alpha V_0} = { \gamma Q_{T}(V_Q) \over Q_0}
\eqno (8)
$$
 The two sides of this equation are plotted in Fig. 5 for different values
 of $\alpha$ and $\sigma$; their intersection points represent our estimate
 for the time-averaged normalized charge trapped in the well during the 
 interaction of the packet with the double barrier. 
 The factor $\gamma$ is fixed by imposing that for $\alpha=0$ the results
 of Fig. 1 are reproduced. As expected it is of the order of unity.
 Fig. 5 predicts correctly the time-averaged obtained from Fig.s 2-4.
 We also notice that for very large $\alpha$ the charge tends to disappear
 due to the smallness of the transmitted amplitude.

We then try to understand the oscillating behavior.
 Let us assume that this phenomenon is due to the competition of two processes:
 {\sl (a)} the filling up of the well by the incoming wave packet and 
 {\sl (b)} the natural decay of the trapped charge.
 For the process {\sl (a)} the time
 scale is of the order of $\hbar / \Gamma_0$ (this estimate appears more 
 reasonable for states narrower in energy than the resonance width).
 For the process {\sl (b)} a reasonable time scale is $\hbar / \Gamma_{V_Q}$,
 where $\Gamma_{V_Q}$ is the energy spread of the function to be integrated
 in Eq. (7) (spectral decomposition of the charge present in the well).
 Oscillations are then expected if a substantial crossover of  $\Gamma_{V_Q}$ 
 and $\Gamma_0$ is realized for the $V_Q$ values reached during the time 
 evolution.
 The analysis of the function 
 $|\widetilde \psi(k,0)|^2 |t_{V_Q}(k)|^2$ shows that $\Gamma_{V_Q}$
 rises from, approximatively, $\Gamma_0 \Gamma_R/ \sqrt{\Gamma_0^2 +
 \Gamma_R^2}$ at $V_Q=0$, to a maximum greater than $\Gamma_0$
 (position and amplitude of the maximum are roughly proportional to
 $\Gamma_0 / \Gamma_R$) and, eventually, decreases to $\Gamma_0$.
 As a consequence when $\Gamma_0 \ll \Gamma_R$, $\Gamma_{V_Q}$ is very  
 close to $\Gamma_0$ and independent of $V_Q$.
 No oscillations are possible in this case for any value of $\alpha$.
 On the other hand,
 when $\Gamma_0 \geq \Gamma_R$, $\Gamma_{V_Q}$ crosses  
 $\Gamma_0$ at a some $V_Q$; oscillations are then realized 
 for a sufficiently high value of $\alpha$.
 This critical value of $\alpha$ increases with the ratio $\Gamma_0 / \Gamma_R$.
 These predictions agree quantitatively with the results of the simulations
 reported above.
 
The effects we have discovered exhibit a considerable stability
 under modifications of the model for example in the direction of a more
 realistic choice of the self-consistent potential.
 A deeper analysis of the problem, together with an
 investigation on the possibility of an experimental study, is matter
 for a forthcoming  publication.
\vglue 2 truecm
\centerline{\sl Acknowledgments}
We thank F. Marchesoni, L. A. Pastur and F. Sacchetti
 for their helpful suggestions and comments. 

\vfill \eject
 
\centerline {\sl References}
\item {[1]} For a review see
 F. Capasso and S. Datta, Phys. Today, {\bf 43} ({\bf 2}), 74, (1990)
 and {\sl Physics of Quantum Electron Devices},
 F. Capasso, ed., Spinger-Verlag, New York, Heidelberg (1990) 
\item {[2]} M. Tsuchiya, T. Matsusue and H. Sakaki, Phys. Rev. Lett.
 {\bf 59}, 2356 (1987); 
 J. F. Young, B. M. Wood, G. C.  Aers, R. L. S. Devine, H. C. Liu,
 D. Landheer, M. Buchanan, A. S. Springthorpe and P. Mandeville,
 Phys. Rev. Lett. {\bf 60}, 2085 (1988); 
 V. S. Goldman, D. C. Tsui and J. E. Cunningham, Phys. Rev. Lett. {\bf 58},
 1256 (1987)
\item {[3]} H. Spohn, Rev. Mod. Phys. {\bf 53}, 569 (1980)
\item {[4]} A. Goldberg, H. M. Schey and J. L. Schwartz,
 Am. J. Phys. {\bf 35}, 177 (1967);
 W. H. Press, B. P. Flannery, S. A. Teukolsky and
 W. T. Vetterling, {\sl Numerical Recipes: the Art of Scientific 
 Computing} (Cambridge University Press, Cambridge 1986)
\item {[5]} S. Collins, D. Lowe and J. R. Barker,
 J. Phys. C  {\bf 20}, 6233 (1987);
 A. Grincwajg, {\sl On the time dependence of double barrier
 resonant tunneling} (Diploma  thesis at the Chalmers University of 
 Technology, G$\ddot o$teborg 1986)
\item {[6]} B. Ricco and M. Ya. Azbel, Phys. Rev. B {\bf 29}, 1970 (1984)
\vfill \eject

\centerline{\sl Figure Captions}
\vglue 1.5 truecm

\item{Fig. 1} Time development of the normalized charge trapped in the well
 in the case of linear Schr$\ddot o$dinger equation, i.e. $\alpha=0$, for
 states with energy spread much wider ($\sigma=110~a_0$ and
 $\Gamma_0=43.2~meV$), of the same order of ($\sigma=825~a_0$ and
 $\Gamma_0=5.8~meV$) and much smaller  ($\sigma=5775~a_0$ and
 $\Gamma_0=0.8~meV$) than the resonance width ($\Gamma_R\simeq 5~meV$).
 An atomic unit of time corresponds to $4.83~10^{-17}$ seconds and the
 Bohr radius $a_0$ is $0.529~\AA$.
\vglue 1.5 truecm

\item{Fig. 2} As Fig. 1 but in the case of effective non linearity in
 the Schr$\ddot o$dinger equation with $\alpha=0.1$.
\vglue 1.5 truecm

\item{Fig. 3} As Fig. 2 but with $\alpha=1$. The case with $\sigma = 5775~a_0$
 is not shown.
\vglue 1.5 truecm

\item{Fig. 4} As Fig. 3 but with $\alpha=10$.
\vglue 1.5 truecm

\item{Fig. 5} Self-consistent estimate of the time-averaged normalized charge
 trapped in the well for different values of the parameters $\alpha$
 and $\sigma$.  
\noindent
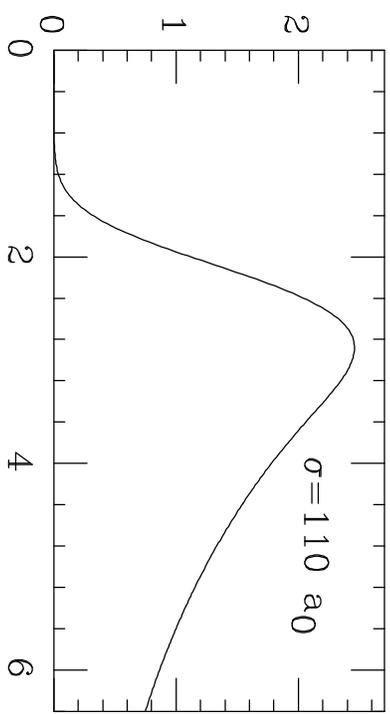
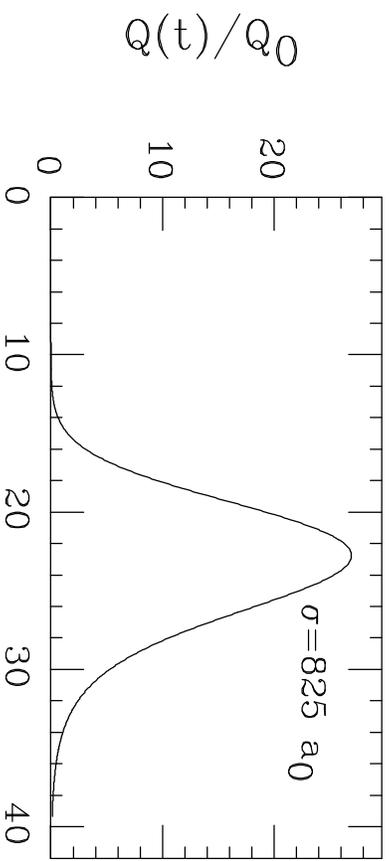
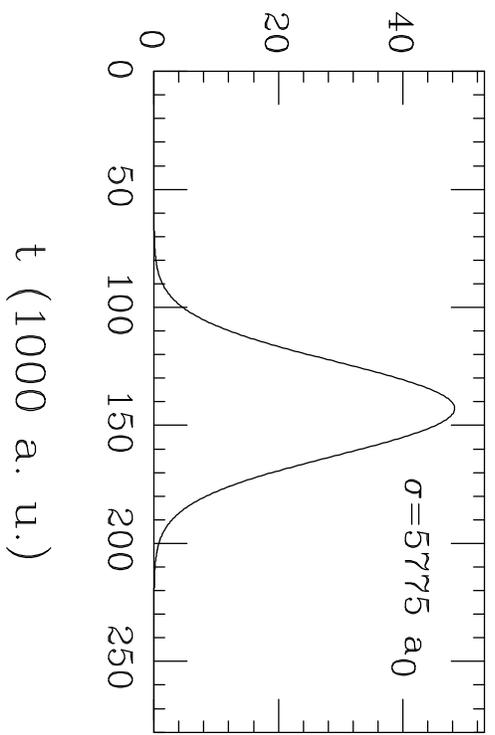
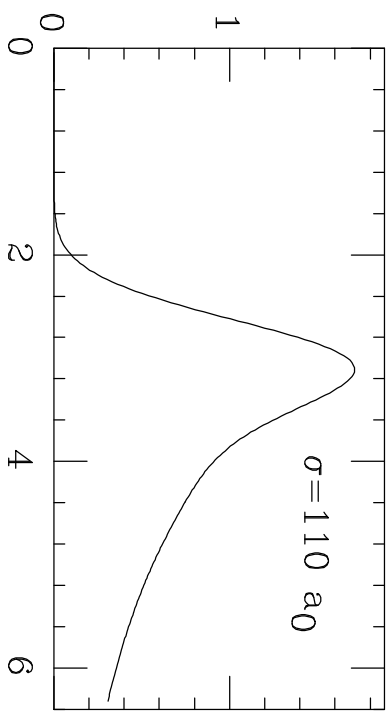
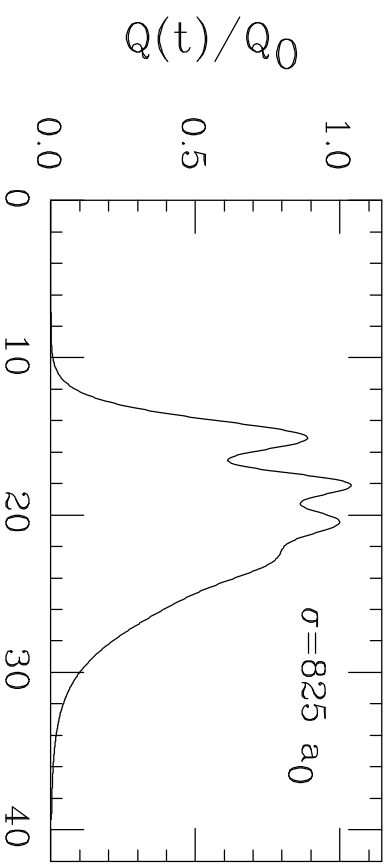
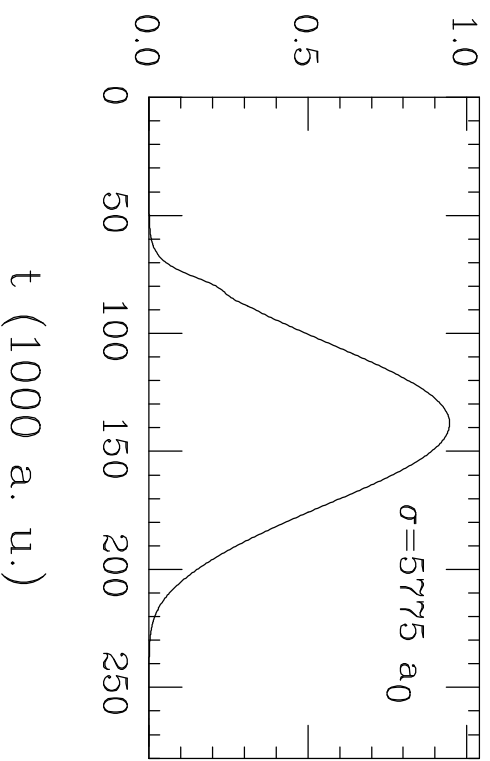
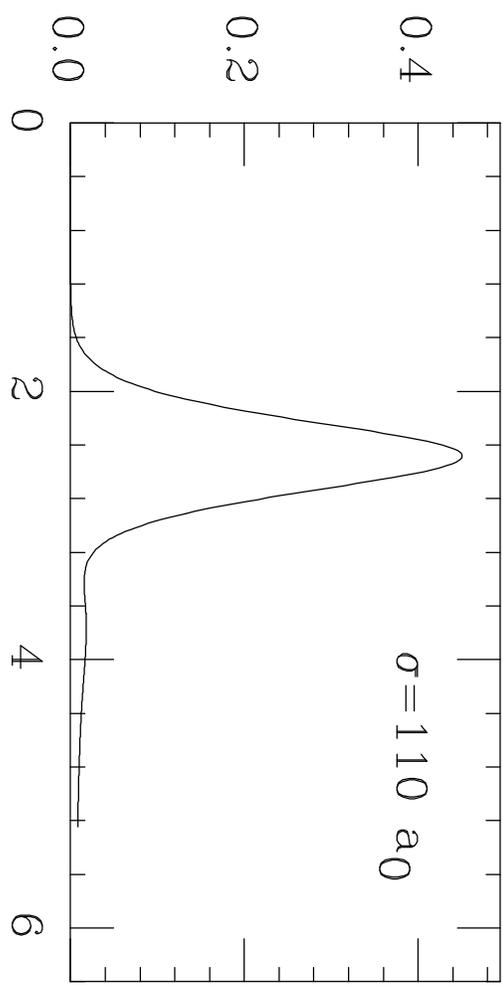
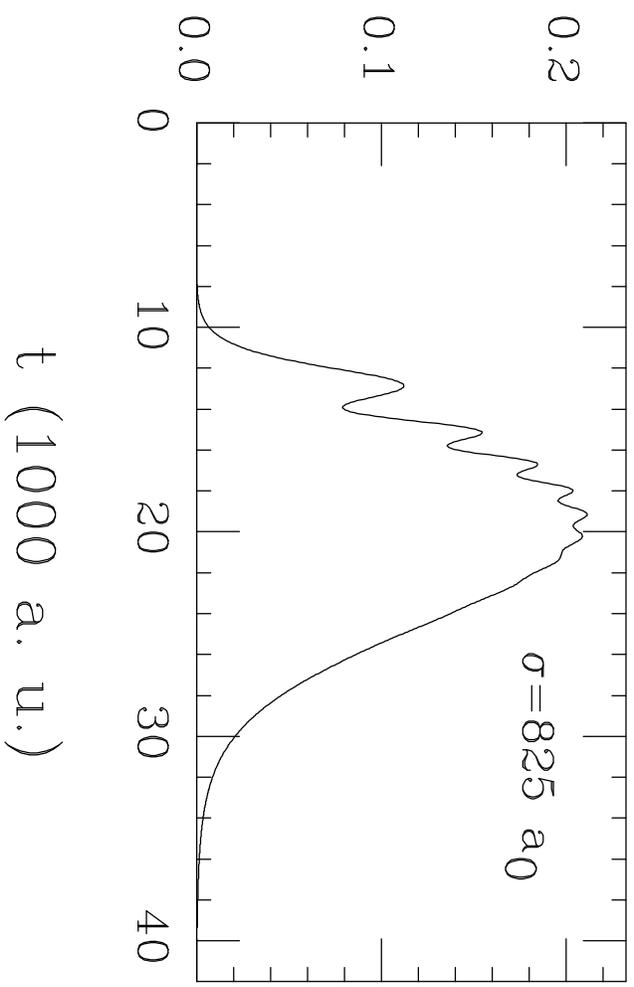
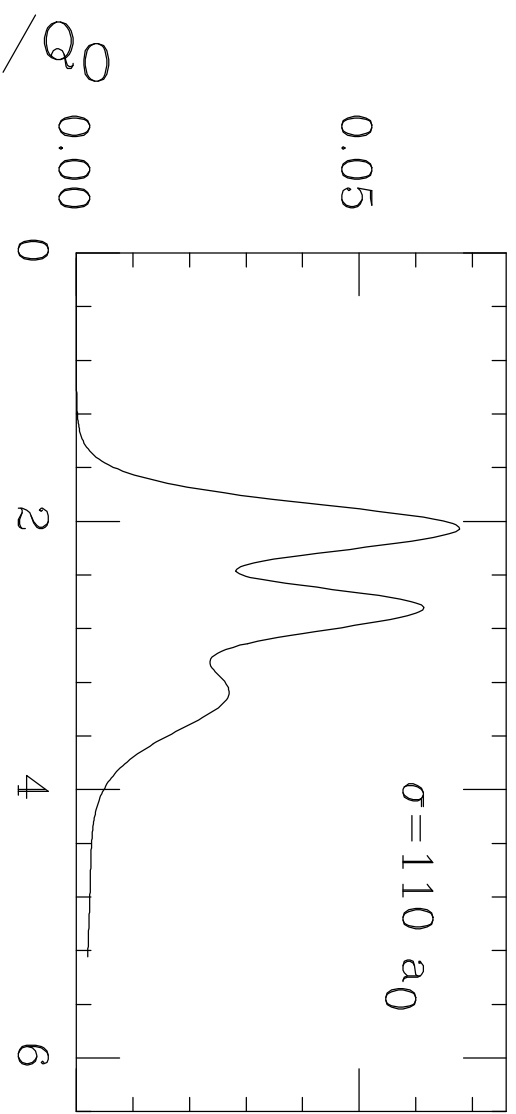
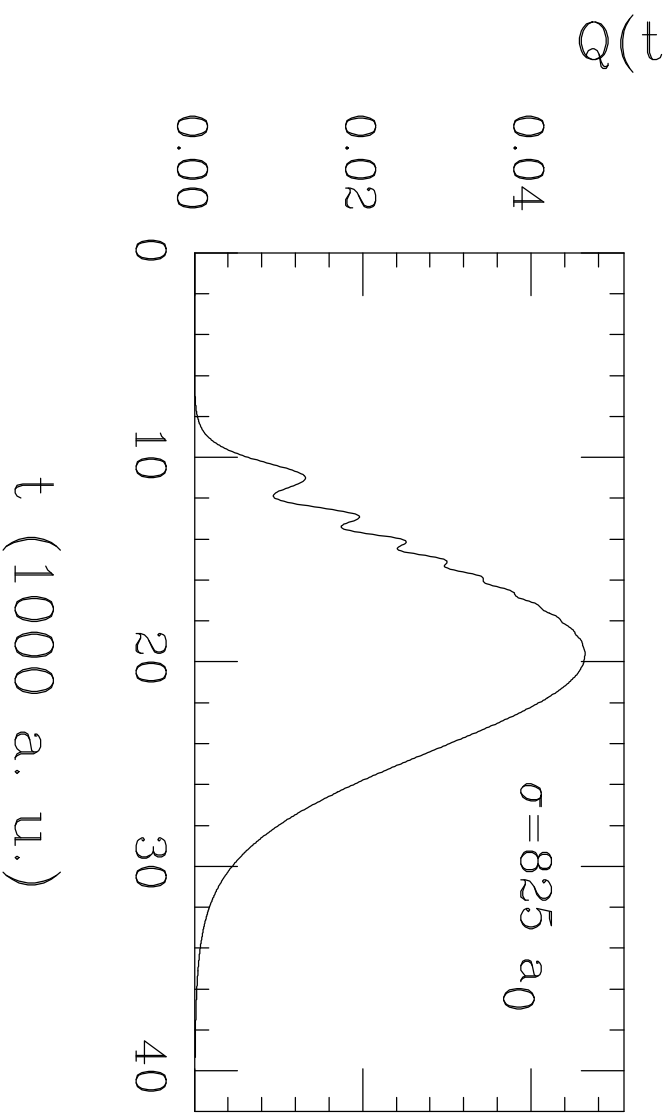
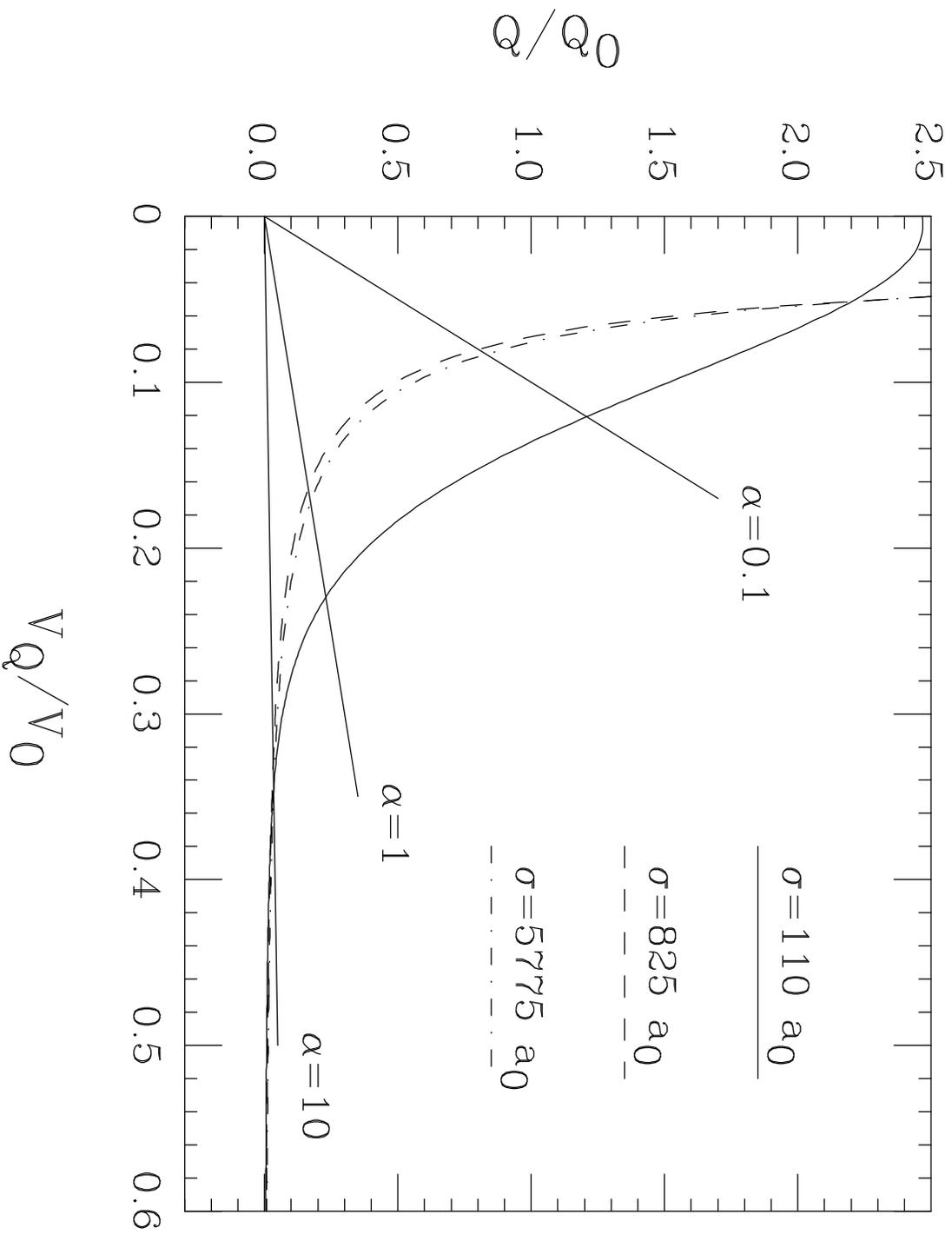
\bye